\documentclass[aps,pra,preprint,superscriptaddress,longbibliography,10pt]{revtex4-1}
\usepackage[utf8]{inputenc}
\usepackage{color, soul}
\usepackage{graphicx}
\usepackage{epstopdf}
\usepackage{siunitx}
\DeclareSIUnit\eV{\eV}
\usepackage{amsmath}
\usepackage{hyperref}
\usepackage{rotating}

\begin{document}

\title{Vorticity-induced flow-focusing leads to bubble entrainment in an inkjet printhead: synchrotron X-ray and volume-of-fluid visualizations}

\author{Maaike Rump}
\thanks{These two authors contributed equally.}
\affiliation{Physics of Fluids group, Max Planck Center Twente for Complex Fluid Dynamics, and J. M. Burgers Centre for Fluid Dynamics, University of Twente, 7500AE Enschede, The Netherlands}
\author{Youssef Saade}
\thanks{These two authors contributed equally.}
\affiliation{Physics of Fluids group, Max Planck Center Twente for Complex Fluid Dynamics, and J. M. Burgers Centre for Fluid Dynamics, University of Twente, 7500AE Enschede, The Netherlands}
\author{Uddalok Sen}
\affiliation{Physics of Fluids group, Max Planck Center Twente for Complex Fluid Dynamics, and J. M. Burgers Centre for Fluid Dynamics, University of Twente, 7500AE Enschede, The Netherlands}
\author{Kamel Fezzaa}
\affiliation{X-ray Science Division, Advanced Photon Source, Argonne National Laboratory, 60439 Illinois, USA}
\author{Michel Versluis}
\affiliation{Physics of Fluids group, Max Planck Center Twente for Complex Fluid Dynamics, and J. M. Burgers Centre for Fluid Dynamics, University of Twente, 7500AE Enschede, The Netherlands}
\author{Detlef Lohse}
\affiliation{Physics of Fluids group, Max Planck Center Twente for Complex Fluid Dynamics, and J. M. Burgers Centre for Fluid Dynamics, University of Twente, 7500AE Enschede, The Netherlands}
\affiliation{Max Planck Institute for Dynamics and Self-Organization, Am Fassberg 17, 37077 Göttingen, Germany}
\author{Tim Segers}
\affiliation{BIOS / Lab-on-a-Chip group, Max Planck Center Twente for Complex Fluid Dynamics, MESA+ Institute for Nanotechnology, University of Twente, 7500AE Enschede, The Netherlands}

\maketitle

\section*{Abstract}
The oscillatory flows present in an inkjet printhead can lead to strong deformations of the air-liquid interface at the nozzle exit. Such deformations may lead to an inward directed air jet with bubble pinch-off and the subsequent entrainment of an air bubble, which is highly detrimental to the stability of inkjet printing. Understanding the mechanisms of bubble entrainment is therefore crucial in improving print stability. In the present work, we use ultrafast X-ray phase-contrast imaging and direct numerical simulations based on the Volume-of-Fluid method to study the mechanisms underlying the bubble entrainment in a piezo-acoustic printhead. We first demonstrate good agreement between experiments and numerics. We then show the different classes of bubble pinch-off obtained in experiments, and that those were also captured numerically. The numerical results are then used to show that the baroclinic torque, which is generated at the gas-liquid interface due to the misalignment of density and pressure gradients, results in a flow-focusing effect that drives the formation of the air jet from which a bubble can pinch-off.

\section{Introduction}

Drop-on-demand inkjet printing is an accurate, non-contact, and highly reproducibly droplet deposition method that can be used to print liquids with a wide range of physical properties~\cite{Wijshoff2010,Derby2010,Hoath2016,Lohse2022}. Today's applications of inkjet printing reach far beyond printing graphics on paper as inkjet technology is now used for the fabrication of electroluminescent displays~\cite{Shimoda2003,Wei2022}, electronic circuits~\cite{Sirringhaus2000,Majee2016,Majee2017}, and biomaterials~\cite{Villar2013,Daly2015,Simaite2016}.

Even though inkjet printing is a highly reproducible droplet deposition technique, the stability of the jetting process can at times be hampered by the entrainment of a bubble~\cite{jeurissen2008effect,jeurissen2009acoustic,jeurissen2011regimes,kim2009effects,lee2009dynamics,vanderbos2011infrared,Lohse2018,Fraters2019Infrared,Lohse2022} . The entrained bubble compromises the channel's incompressibility and it thereby disturbs the acoustics in the printhead and can even stop the entire jetting process~\cite{dejong2006entrapmentacoustic,dejong2006entrapmentfluids,Fraters2019Nucleation}. As applications of inkjet printing rely on the stability and reproducibility of the droplet formation process, it is crucial to understand the physical mechanisms of the bubble entrainment process.

Several bubble entrainment mechanisms have been identified. First, an ink layer or particles on the nozzle plate can result in bubble entrainment upon their interaction with the ink jet during its ejection from the nozzle~\cite{dejong2006entrapmentacoustic,Li2019}. Second, bubbles can nucleate on microparticles present in the ink due to cavitation during the rarefactional pressure wave or upon their direct interaction with the oscillating meniscus~\cite{Fraters2019Nucleation}. Third, bubbles can be entrained through direct bubble pinch-off from the oscillating meniscus, e.g. as a result of asymmetric meniscus oscillations~\cite{vdMeulen2020,Fraters2021Oscillations}. Bubble entrainment mechanisms are typically stochastic, where the probability of bubble entrainment increases exponentially with the driving amplitude~\cite{Fraters2019Nucleation}. At high driving amplitudes, however, bubble entrainment due to meniscus instabilities can become deterministic, as we have shown before using an optically transparent printhead~\cite{Fraters2021Oscillations}. Here, the high-frequency flow oscillations in the printhead can result in the formation of an air jet on the retracted concave meniscus~\cite{Fraters2021Oscillations}. The tip of the air jet can pinch-off, resulting in bubble entrainment. We used inviscid irrotational boundary integral (BI) simulations to identify flow-focusing as the physical mechanism that drives this meniscus instability, leading to deterministic bubble pinch-off. Geometrical flow-focusing is the process where, when a pressure wave arrives at a concave part of the meniscus, the inhomogeneous pressure gradient field along the meniscus leads to the formation of a jet~\cite{Antkowiak2007,Kiyama2016,Tagawa2012,Delrot2016,Gordillo2020}. However, since we used inviscid and irrotational BI simulations, a direct comparison between experiment and numerics is not feasible as viscous effects are ignored. Therefore, open questions have remained as to the role of viscosity and vorticity on the bubble pinch-off mechanism driven by flow focusing~\cite{Fraters2021Oscillations}. A direct comparison between experiment and numerics was further complicated by the refractive index mismatch between ink and air, which masks certain regions of the curved meniscus. As such, only part of the meniscus oscillations could be visualized in high-speed microscopy.

In this paper, we will overcome these limitations: On the experimental side, by using X-rays for imaging rather than light in the visible range, and on the numerical side by using a Volume-of-Fluid (VoF) method rather than BI. X-rays do not suffer as much from refraction at a curved interface (see Fig.~\ref{fig:setupX}d) as visible light rays, due to the short wavelength of the photons. High-energy X-rays are typically produced in a synchrotron. Ultrafast X-ray phase-contrast imaging has been previously used to study droplet coalescence~\cite{Fezzaa2008}, drop impact on liquid pools~\cite{Zhang2012,Lee2015} and substrates~\cite{Lee2012,Lee2020}, and droplet formation in drop-on-demand inkjet printing~\cite{Parab2019}. The frame rate in the synchrotron is limited to \SI{270}{\kilo\hertz} by the roundtrip of electrons in the storage ring~\cite{StorageRing}. However, stroboscopic high-speed imaging can be performed using multiple interleaved high-speed recordings to reach an equivalent frame rate of 2 million frames per second. A prerequisite for stroboscopic imaging is a fully deterministic and reproducible process, which turned out from earlier experiments in the case for the bubble pinch-off phenomenon studied here~\cite{Fraters2021Oscillations}.

In the present study, we investigate the physics governing the geometric flow-focusing mechanism of bubble entrainment in a piezoacoustic printhead. The detailed meniscus oscillations are visualized using ultrafast phase contrast X-ray imaging, which allowed us to not only capture the complete interface dynamics, but also to extract the displaced volume, and thereby the flow rate. The experimentally obtained flow rate provides the input for our direct numerical simulations (DNS) performed using the VoF method, which was previously successfully used in studies involving drops and bubbles~\cite{Fuster2009,Zhang2015,Thoraval2016,Conto2019,Ganan2021}. We use the simulations to unravel the intricate local flows during the bubble entrainment process, which reveal the origin of the geometric flow-focusing mechanism in our printhead. We furthermore use the numerics to study the influence of viscosity and vorticity on the bubble pinch-off process.

The work is organized as follows: In \autoref{sec:Exp} and \autoref{sec:Num}, the experimental and numerical methods are described, respectively. Then, in \autoref{sec:Validation}, the comparison between the experiments and numerics is discussed. The different observed bubble pinch-off classes are presented in \autoref{sec:Types}. In \autoref{sec:Mech}, we describe our findings on the mechanism of bubble entrainment via geometrical flow-focusing. The work ends with the conclusions in \autoref{sec:Conclusion}.

\section{Ultrafast X-ray phase contrast imaging}\label{sec:Exp}
\subsection{Set-up and procedure}

\begin{figure}[ht]
    \centering
    \includegraphics[width = .9\textwidth]{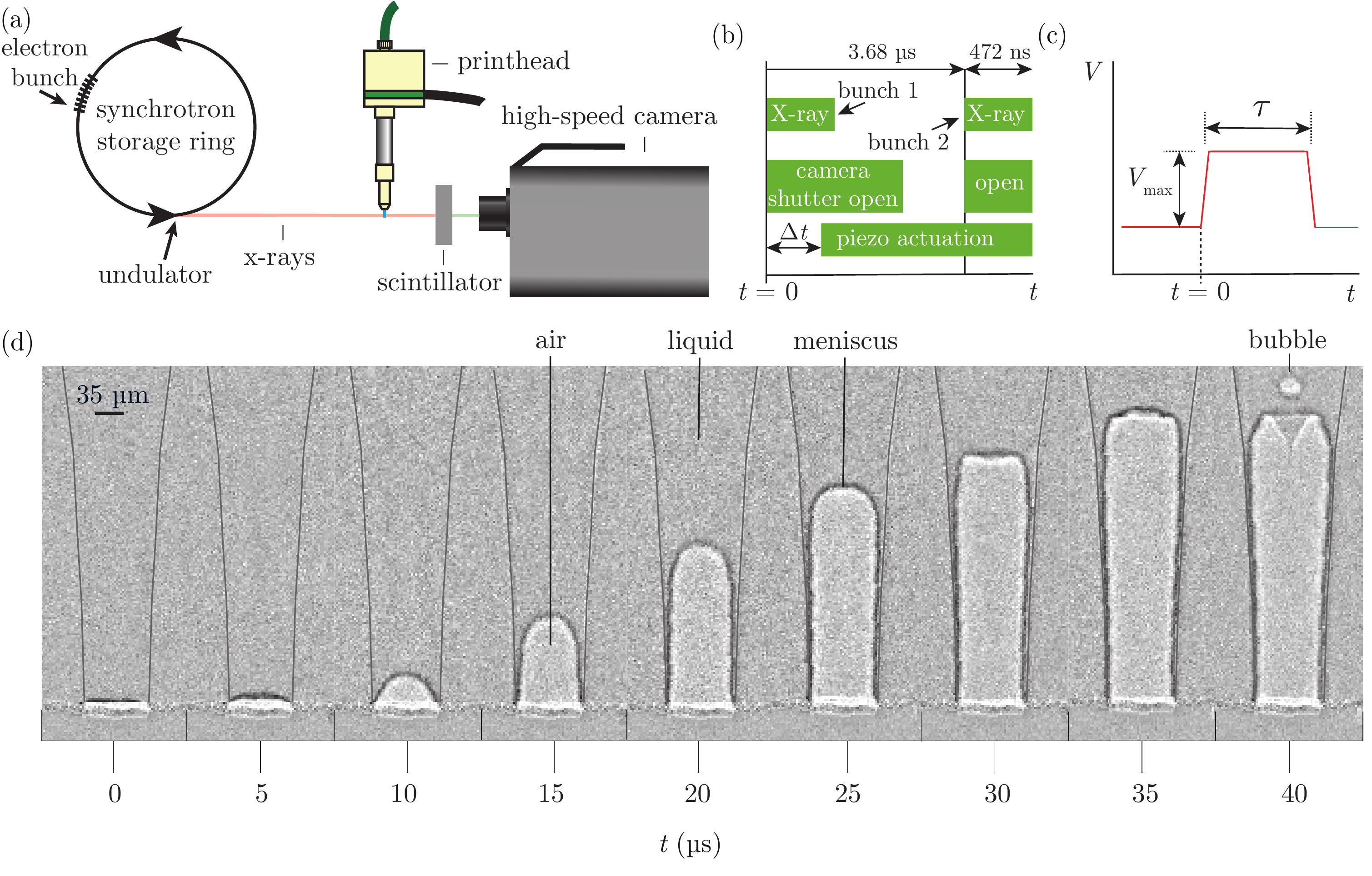}
    \caption{(a) Schematic of the experimental setup. (b) $t=0$ is when the first \SI{472}{\nano\second} X-ray pulse illuminates the printhead and the camera is triggered. The piezo in the printhead is triggered after a set delay $\Delta t$, starting at \SI{0}{\micro\second}. After \SI{3.68}{\micro\second} the next X-ray pulse illuminates the set-up for the second frame in the recording. This process continues for 200 frames. The next recording has a delay $\Delta t$ of \SI{0.5}{\micro\second} for the piezo actuation. This continues in steps of \SI{0.5}{\micro\second} until a maximum $\Delta t$ of \SI{3.5}{\micro\second}. (c) The trapezoidal piezo driving pulse. (d) Typical bubble entrainment phenomenon for a driving pulse with $V_{max} =$ \SI{140}{\volt} and $\tau =$ \SI{32}{\micro\second}. The edge of the nozzle is marked by the gray lines.}
    \label{fig:setupX}
\end{figure}

A schematic of the experimental imaging setup is shown in Fig.~\ref{fig:setupX}a. The experiments were performed at the 32-ID undulator beamline at the Advanced Photon Source of the Argonne National Laboratory with the storage ring in hybrid mode. In this hybrid mode, 84\% of all electrons stored in the synchrotron storage ring are collected in a bunch. The electrons produce X-rays as they are forced through an undulating trajectory by magnetic structures. These X-rays can be released unto the set-up by opening a shutter. The resulting X-ray illumination pulse has a duration of~\SI{472}{\nano\second} and a peak irradiance of $10^{14}$~ph/s/mm$^2$/0.1$\%$bw~\cite{Fezzaa2008}. The shutter releasing the X-rays, the printhead, and the high-speed camera (Photron Fastcam SA-Z) were synchronized to the passing of the electron bunch in the storage ring using delay generators (Standford DG535) that were triggered using the radio-frequency signal generated by the passing of the electron bunch. The time between two subsequent X-ray illumination pulses is set by the round-trip time of the electron bunch of \SI{3.68}{\micro\second} in the storage ring. The interframe time of the high-speed camera was therefore also set to \SI{3.68}{\micro\second} (271.553~frames/s) such that every camera frame was exposed by a single X-ray illumination pulse, see Fig.~\ref{fig:setupX}b. The X-rays were converted into visible wavelength photons using a scintillator plate (LuAG:Ce, decay time \SI{50}{\nano\second}), which were captured by the high-speed camera and a 10$\times$ imaging objective (Mitutoyo Plan Apo Infinity Corrected Long WD, Edmund Optics) at a spatial resolution of \SI{2}{\micro\meter}/pixel. Figure~\ref{fig:setupX}d shows the meniscus retracting into the nozzle. Around $t =$ \SI{30}{\micro\second}, the concave meniscus deforms and an instability develops, which pinches of a bubble. Each experiment was repeated 9 times, where for every repeat the delay time ($\Delta$t) was increased by \SI{0.5}{\micro\second} for the start of the piezo driving pulse, as illustrated in Fig.~\ref{fig:setupX}b. The nine high-speed recordings were then interleaved to obtain a single high-speed recording at an effective frame rate of 2 million frames/s (\SI{500}{\nano\second} interframe time).

The employed single-nozzle printhead (Autodrop Pipette, Ad-K-501 and AD-H-501, Microdrop Technologies GmbH) has a nozzle diameter of \SI{70}{\micro\meter}~\cite{Dijksman1984,Dijksman1998}. The printhead was driven by a rectangular pulse (Fig.~\ref{fig:setupX}c) with rise and fall times of \SI{0.2}{\micro\second}. The piezo driving pulse was generated by an arbitrary waveform generator (Agilent 33220A) and amplified by a broadband amplifier (Falco System WMA-300). The pulse width $\tau$ was varied between \SI{30}{\micro\second} and \SI{36}{\micro\second} and the driving amplitude $V_{max}$ was varied between \SI{70}{\volt} and \SI{150}{\volt}. In all experiments, Milli-Q water was used as the model ink. An underpressure was applied by manually adjusting the piston of the syringe, which was connected to the liquid supply line of the printhead, to prevent the water from dripping outward under the influence of gravity.

\subsection{Image processing}

The grayscale images obtained from the experiments appear uniform in brightness, except at the interfaces (Fig.~\ref{fig:setupX}d). This is the result of the relatively high X-ray energy, which essentially overexposes the images. However, due to the coherence of the X-ray beam, the X-rays are refracted at the interfaces and form a pair of a dark and a bright fringe at the location of the interface as the waves propagate from the printhead to the camera~\cite{Fezzaa2008}. The distance between the setup and the camera was optimized (approximately \SI{0.4}{\meter}) to achieve the highest contrast of said interface-fringes, where the gas-liquid interface is located between the dark and the bright fringe~\cite{Wilkins1996,Mayo2012}. In Fig.~\ref{fig:setupX}d, it is shown that the dark part of the fringe is located at the liquid side while the bright part is located at the air side.

\begin{figure}[ht]
    \centering
    \includegraphics[width = \textwidth]{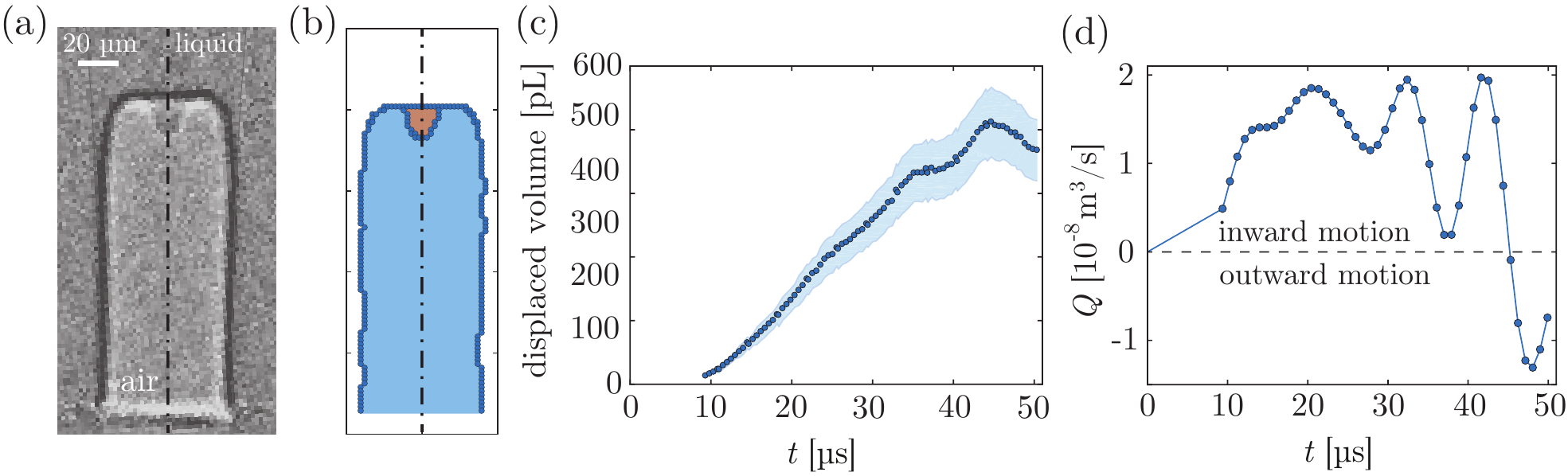}
    \caption{(a) Experimental snapshot where the black dash-dotted line is the axis of symmetry of the nozzle. (b) Edge detection result of the snapshot shown in (a). The total volume minus the volume in the orange area results in the volume of the blue region. (c) Extracted volume over time for $V_{max}=$ \SI{70}{\volt} and $\tau=$ \SI{34}{\micro\second}. (c) Flow rate ($Q$) as determined from the volume curve in (b). The experimentally obtained $Q(t)$ was used as an input to the numerical simulations shown in Fig.~\ref{fig:comparison}.}
    \label{fig:extracting_flowrate}
\end{figure}

The analysis of the images was performed in Matlab using the Canny edge detection method. An example of the image processing result on the frame shown in Fig.~\ref{fig:extracting_flowrate}a is presented in Fig.~\ref{fig:extracting_flowrate}b. To extract the volume of air in the nozzle from the images, we use the axis of symmetry of the nozzle being in the middle between the two nozzle side walls (black dash-dotted line) and then extract the distance of each edge point (the blue datapoints in Fig.~\ref{fig:extracting_flowrate}b) to the axis of symmetry. Two consecutive radii were then used to calculate the volume of a truncated cone, by assuming cylindrical symmetry~\cite{vanderbos2014}. The total volume is the summation of all truncated cones along the center line, indicated by the blue region. When an outward-directed liquid jet was formed into the air-filled cavity, as in Fig.~\ref{fig:extracting_flowrate}a, the volume of the liquid jet was found in the same way (orange area) and then subtracted from the total volume to obtain the volume of air. The resulting displaced volume of air over time is shown in Fig.~\ref{fig:extracting_flowrate}c. The meniscus position could not be analyzed for times before \SI{9}{\micro\second} as it was optically-overlapping with the glass-air interface. Therefore, the experimentally-measured displaced volume was linearly interpolated between $t=0$ and $t=\SI{9}{\micro\second}$, from zero to the first measured displaced volume. The curve in Fig.~\ref{fig:extracting_flowrate}c was smoothed and resampled (Matlab's smoothing spline with smoothing parameter of 0.7), and its derivative provides the volume flow rate $Q$ (Fig.~\ref{fig:extracting_flowrate}d), which was used as an input for the numerical simulations.

\section{Numerical method}\label{sec:Num}
\subsection{Governing equations}
To gain insight into the mechanism of bubble entrainment, numerical simulations were performed with the free software program \textsc{basilisk}~\cite{Popinet2015}, which solves partial differential equations on an adaptive Cartesian mesh (quad/octree discretization) employing the Volume-of-Fluid method. The governing equations of our system are:

\begin{equation}
	\nabla\cdot\boldsymbol{u} = 0,
	\label{eq::continuity}
\end{equation}
\begin{equation}
	\rho\left(\frac{\partial\boldsymbol{u}}{\partial t} + \boldsymbol{u}\cdot\nabla\boldsymbol{u}\right) = -\nabla p + \nabla\cdot\left[\mu\left(\nabla\boldsymbol{u} + \nabla\boldsymbol{u}^T\right)\right] + \sigma\kappa\delta_S\boldsymbol{n},
	\label{eq::momentum}
\end{equation}
reflecting the conservation of mass and momentum in an incompressible formulation, respectively. In the above equations, $\boldsymbol{u}$ is the velocity vector field, $p$ the pressure, $\rho$  the density, and $\mu$ the dynamic viscosity. The last term in Eq.~(\ref{eq::momentum}) describes capillary forces, modelled as continuous surface forces (CSF)~\cite{Brackbill1992}, where $\sigma$ is the surface tension coefficient, $\kappa$ the curvature of the interface, $\delta_S$ a characteristic function defined as 1 at the interface and 0 elsewhere, and $\boldsymbol{n}$ the normal vector to the interface.

These equations were solved via a finite volume discretization. The numerical scheme is detailed in~\cite{Popinet2003,Popinet2009,Lagree2011}. To account for multiple phases, a geometric Volume-of-Fluid (VoF) method was employed~\cite{Scardovelli1999}, in which the interface has a sharp representation. In our two-phase framework, a volume fraction $c$ delimits both phases. It is equal to 1 in water, to 0 in air, and to values in between in cells containing an interfacial segment. The volume fraction is advected with the local velocity field, thus obeying the following equation:
\begin{equation}
	\frac{\partial c}{\partial t} + \boldsymbol{u}\cdot\nabla c = 0.
\end{equation}
The flow was modeled in a one-fluid formulation, with one continuous velocity field. Jumps of the local fluid properties, such as density and viscosity, were allowed across the interface, where they were defined as arithmetic averages, weighted by the volume fraction: $\{\rho,\mu\}=c\{\rho_1,\mu_1\}+(1-c)\{\rho_2,\mu_2\}$, where 1 and 2 are subscripts denoting water and air, respectively. In the presence of capillary forces, jumps in pressure across the interface are also taken into account.

\subsection{Numerical setup}
\begin{figure}[ht]
	\centering
	\includegraphics[width = \textwidth]{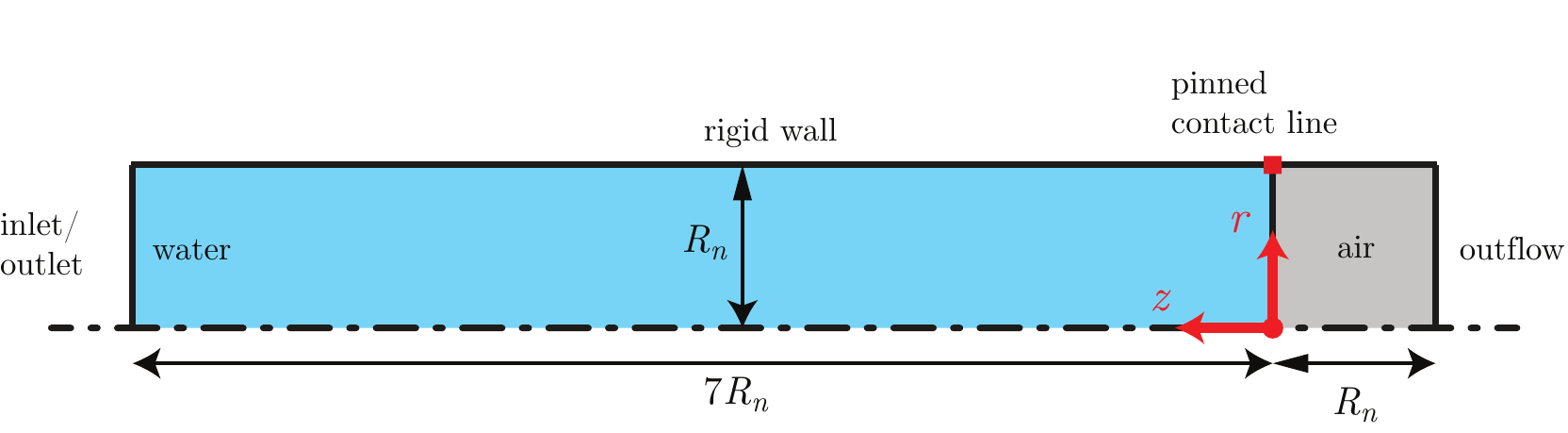}
	\caption{Schematic of the axisymmetric numerical domain (to scale), with the boundary conditions specified. The bottom boundary is the axis of symmetry, depicted by the dot-dashed line.}
	\label{fig:schematic}
\end{figure}

Figure~\ref{fig:schematic} depicts the numerical domain in the initial quiescent configuration. The experimental nozzle is tapered, with its local radius being a function of the axial coordinate $z$. However, in our numerics, this tapering is neglected. Therefore, the nozzle is modeled in axisymmetric cylindrical coordinates with constant radius $R_n$~=~\SI{35}{\micro\meter}, corresponding to the radius of the nozzle exit. We compare our numerical results with experiments performed at a low driving voltage, where the liquid does not retract much into the nozzle, and where the effect of the tapering of the experimental nozzle is thus expected to be limited. It is to be noted that the tapering can also result in the curving of an accelerating liquid-air interface \cite{villermaux-2010-jfm}. However, in the present work, we focus on the interface dynamics of a driven system having a complex flow rate (determined from the experiments) set as the initial conditions. Thus we only focus on cases corresponding to lower driving voltages, where the effect of the tapering is expected to be minimal. The numerical domain is rectangular, with dimensions $\left[8:1\right]R_n$. The origin is set at the nozzle exit, indicated by the polar coordinates' unit vectors in Fig.~\ref{fig:schematic}. Liquid water initially occupies the region of the domain where $z\geq0$. Air occupies the remaining region where $z<0$.

The bottom boundary is the axis of symmetry, depicted by the dot-dashed line. The top boundary has the conditions of a rigid wall, i.e. $\boldsymbol{u}=\boldsymbol{0}$, $\partial p/\partial r=0$. Generally, the VoF method allows accurate control over the contact line behavior via the setting of well defined boundary conditions on the volume fraction $c$~\cite{Afkhami2018}. For example, one can fix a specific contact angle throughout the simulation or pin the contact line at specific locations. In the present work, we pin the contact line at the nozzle exit, analogously to experimental observations (Fig.~\ref{fig:setupX}d), by setting a condition on the volume fraction,
\begin{equation}
	c(r = R_n) =
	\begin{cases}
		1, & \text{if } z\geq0,\\
		0, & \text{otherwise.}
	\end{cases}
\end{equation}
The right boundary has outflow conditions, where the pressure is set to a gauge constant $p=0$, and $\partial\boldsymbol{u}/\partial z=\boldsymbol{0}$. Finally, the flow is driven at the left boundary, having the conditions of an inlet/outlet. Although the flow is continuously transient, and not fully developed, a Poiseuille velocity profile is prescribed at the left boundary for it to respect the no-slip boundary condition at the rigid wall,
\begin{equation}
	u_z(r,t) = -\frac{2Q(t)}{\pi R_n^2}\left[1 - \left(\frac{r}{R_n}\right)^2\right],
\end{equation}
 where $u_z$ is the axial component of the velocity. This parabolic profile is defined as $\int u_z(r,t)dA=-Q(t)$, where $Q(t)$ is the instantaneous flow rate through the nozzle as measured from the experiment (Fig.~\ref{fig:extracting_flowrate}c) and $dA$ a differential cross-sectional area. The pressure is set to a zero Neumann boundary condition in the normal direction, i.e. $\partial p/\partial z=0$. 

Adaptive mesh refinement was employed in the numerical simulations, based on the error of the volume fraction $c$ and the velocity field $\boldsymbol{u}$. Therefore, the grid was refined at the interface and in the regions of interest. The smallest grid size attained in the simulations was $\Delta=R_n/128$, which was found sufficient to be for our current purposes.

\section{Results \& Discussion}

\subsection{Phenomenology and numerical validation}\label{sec:Validation}

\begin{figure}[ht]
    \centering
    \includegraphics[width = .9\textwidth]{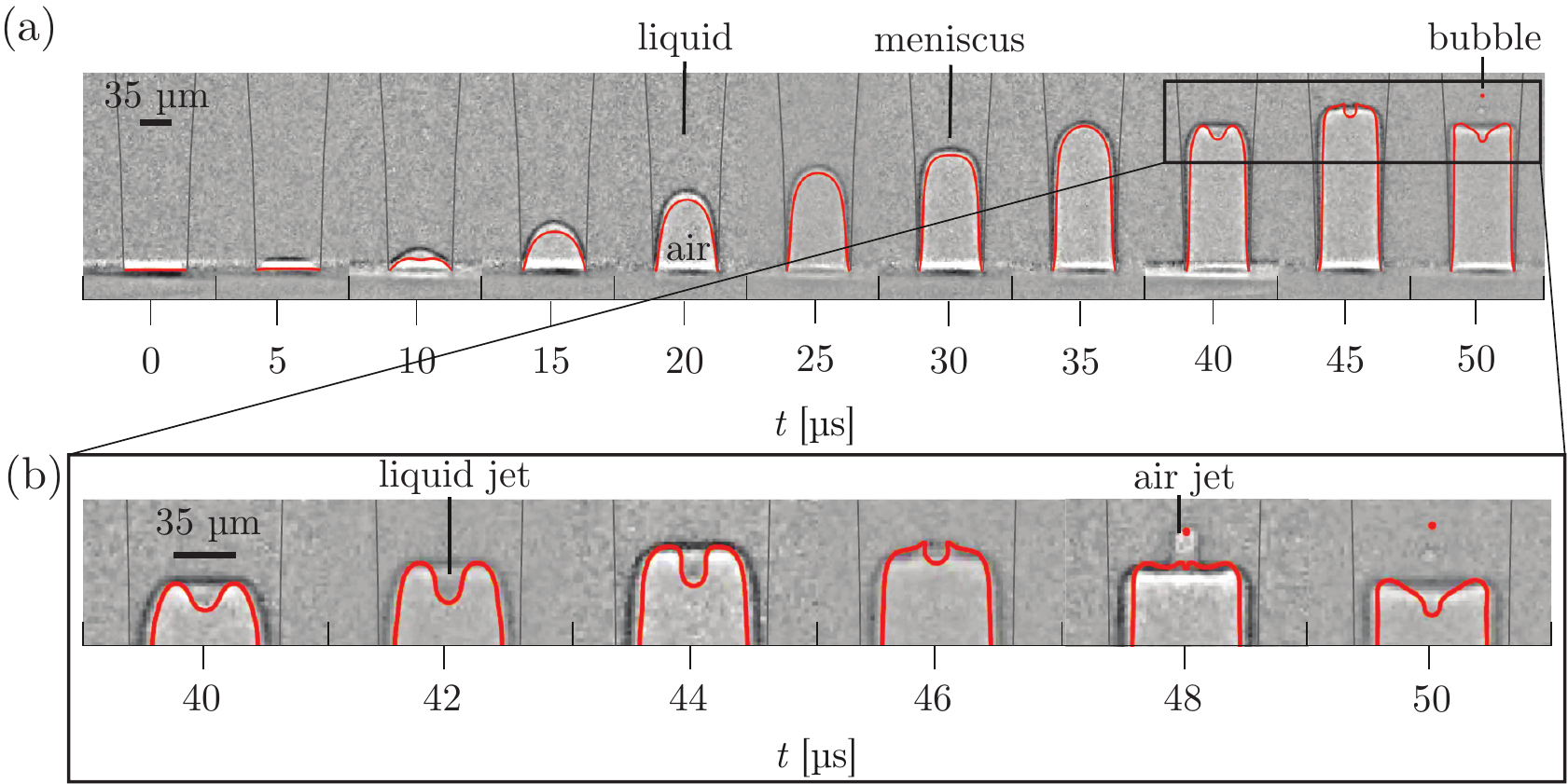}
    \caption{(a) Meniscus position during different time instants of the bubble entrainment process as observed in the experiment (grayscale images) and the numerical simulations (red curves) for $V_{max}=$ \SI{70}{\volt} and $\tau=$ \SI{34}{\micro\second}. (b) A zoomed-in image of the same bubble entrainment phenomenon. The outline of the nozzle is highlighted with gray lines.}
    \label{fig:comparison}
\end{figure}

Figure~\ref{fig:comparison}a shows a typical example of a bubble entrainment process inside the nozzle of the piezoacoustic printhead as observed using X-ray imaging. The corresponding numerical simulations are depicted by the solid red curves. A movie of the process can be found in the supplementary material (SM1~\cite{SM}). Good agreement is observed between the experimental and numerical results, with only slight discrepancies due to the missing tapering in the numerical simulations. The motion of the meniscus near the moment of bubble pinch-off is shown in more detail in Fig.~\ref{fig:comparison}b. The bubble entrainment process can be described as follows: First, the meniscus retracts inward ($10-$\SI{35}{\micro\second}), thereby becoming concave in shape after which it is accelerated outwards (\SI{40}{\micro\second}), producing a liquid jet in the center (Fig.~\ref{fig:comparison}b). Subsequently, another outward acceleration moves the outer region of the meniscus outward ($46-$\SI{48}{\micro\second}) whereas the center of the meniscus stays behind, causing the central jet to recoil, thereby forming an inward air jet in the center (\SI{48}{\micro\second}, Fig.~\ref{fig:comparison}b). When this air jet pinches off, a bubble is entrained (\SI{50}{\micro\second}), similar to the bubble entrainment process described before in~\cite{Fraters2021Oscillations}.

\subsection{Bubble pinch-off classes}\label{sec:Types}
\begin{sidewaysfigure}
	\centering
	\includegraphics[width = \textwidth]{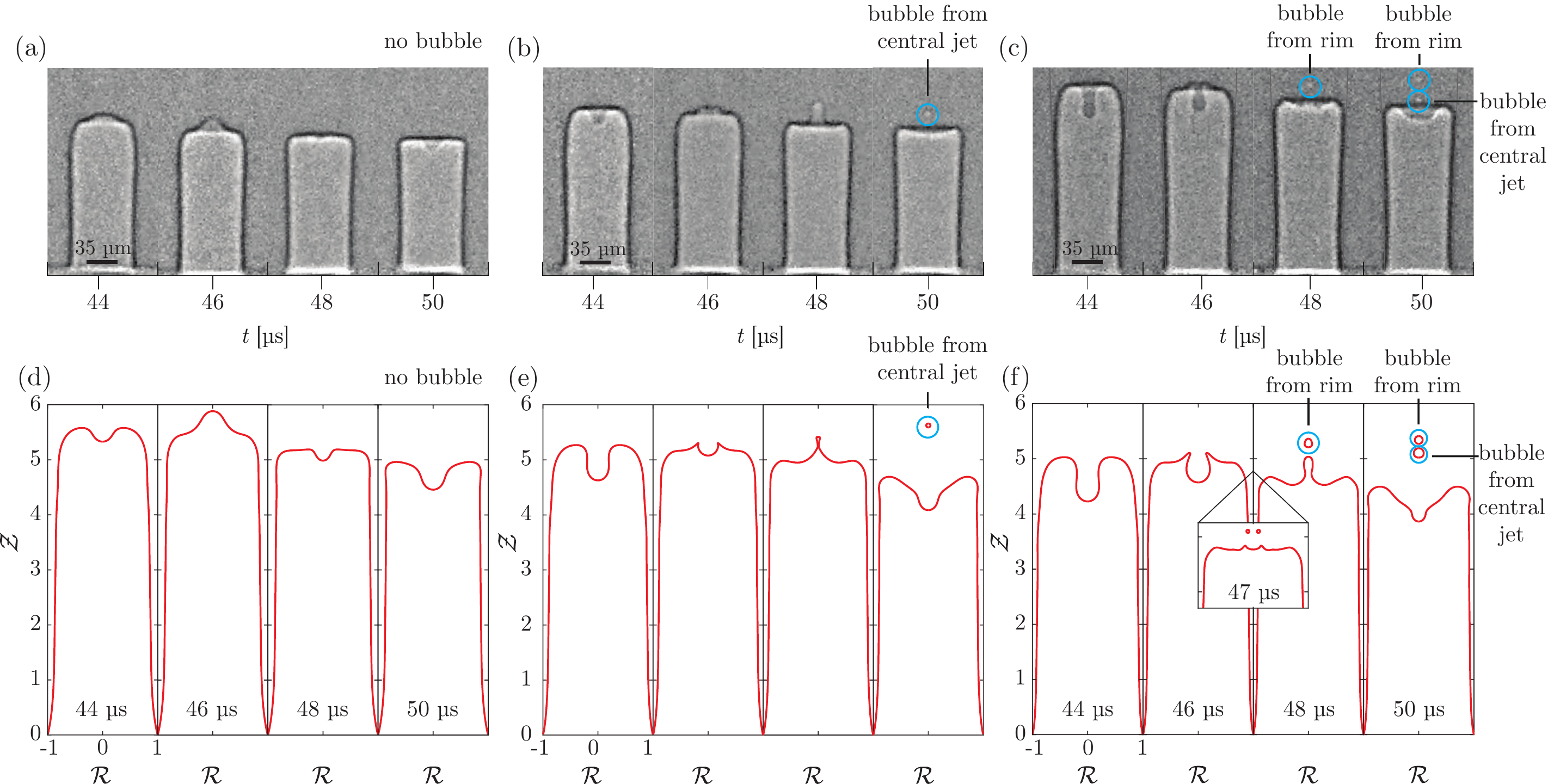}
	\caption{(a) No bubble entrainment in the experiment with $V_{max}=$ \SI{70}{\volt} and $\tau=$ \SI{32}{\micro\second}. (c) One bubble is entrained in the experiment at $V_{max}=$ \SI{70}{\volt} and $\tau=$ \SI{34}{\micro\second}, which is the same experiment as in Fig.~\ref{fig:comparison}. (c) Two bubbles are entrained in the experiment at $V_{max}=$ \SI{80}{\volt} and $\tau=$ \SI{34}{\micro\second}. (d) No bubble entrainment in the numerics with a viscosity of \SI{2}{\milli\pascal\cdot\second}. (e) One bubble is entrained in the numerics with a viscosity of \SI{1}{\milli\pascal\cdot\second}, which is the same simulation as in Fig.~\ref{fig:comparison}. (f) Two bubbles are entrained in the numerics with a viscosity of \SI{0.5}{\milli\pascal\cdot\second}. In all the numerical results, the nozzle dimensions are non-dimensionalized by the nozzle radius $R_n=$ \SI{35}{\micro\meter}.}
	\label{fig:types}
\end{sidewaysfigure}

In the search for different bubble pinch-off regimes, several parameter scans were performed. Note that the aim of this section is to provide a phenomenological, rather than a one-to-one quantitative comparison between the experiments and the numerics. Experimentally, the driving amplitude $V_{max}$ and pulse width $\tau$ were varied while the liquid properties were kept constant. For high driving amplitudes, the meniscus retracts further into the nozzle whose tapering starts to have a significant effect on the dynamics, leading to large discrepancies between the experiments and the numerics. Therefore, to numerically capture the different classes of bubble entrainment that were experimentally observed, the effect of inertia on the meniscus oscillations is replicated by changing the liquid viscosity, while the flow rate boundary condition was kept constant (constant low amplitude driving). The different bubble entrainment classes are depicted in Fig. \ref{fig:types}.

We start with the reference case, depicted in Figs.~\ref{fig:types}b and e. It shows the pinch-off of a spherical bubble from the central air jet. No bubble is entrained when the pulse width is decreased in the experiment (Fig.~\ref{fig:types}a), or when the viscosity is increased in the numerics (Fig.~\ref{fig:types}d). On the other hand, when in the experiment the driving amplitude is increased, or when the viscosity is decreased in the numerics, two bubbles are entrained as shown in Figs.~\ref{fig:types}c and \ref{fig:types}f, respectively. The first bubble appears when the central liquid jet is present. At this moment, the rim at the base of the liquid jet pinches off to form a toroidal bubble that quickly becomes spherical due to surface tension (to minimize its surface energy), as also illustrated in the inset of Fig.~\ref{fig:types}f. Toroidal bubble pinch-off was also mentioned in ref.~\cite{Fraters2021Oscillations}. The second entrained bubble again pinches-off spherically from a central jet, similar to the scenario shown in Figs.~\ref{fig:types}b and e. It is evident from Fig.~\ref{fig:types} that even though the parameters varied in the experiments (voltage and pulse width) and the numerics (liquid viscosity) are different, the numerical simulations can still capture the meniscus motion and the bubble entrainment (both from the central jet and from the rim of the liquid jet) with a high degree of accuracy.

\subsection{Physical mechanism of bubble pinch-off}\label{sec:Mech}
\begin{figure}[ht]
    \centering
    \includegraphics[width = \textwidth]{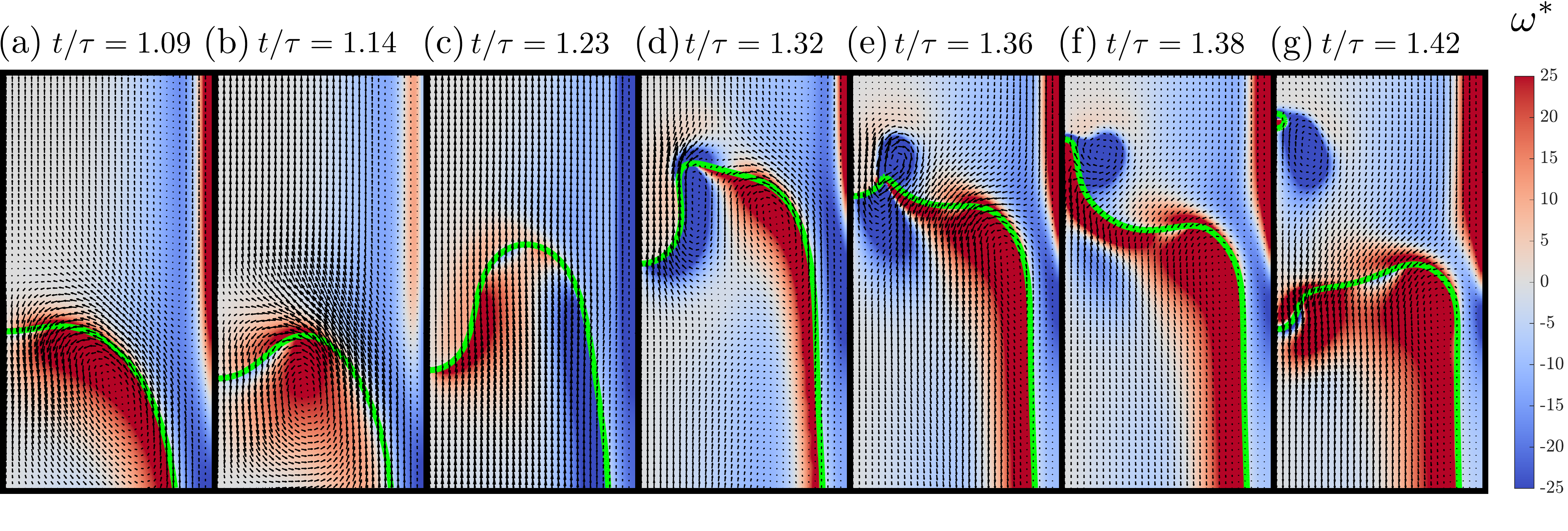}
    \caption{Snapshots from numerical simulations of the meniscus oscillations leading to bubble entrainment. The velocity is depicted as a vector field, and the dimensionless vorticity $\omega^* = \omega\tau$ is illustrated in the background as a color code. In each snapshot, the interface is drawn in green. This is the $V_{max}=$ \SI{70}{\volt} and $\tau =$ \SI{34}{\micro\second} case, where the fluid properties are as those in the experiment.}
    \label{fig:vecotrs}
\end{figure}
In this section, we identify the mechanism of bubble entrainment using the information provided by the numerical simulations. The traditional picture of the process leading to bubble pinch-off is one driven by surface tension, and the focusing of capillary waves at the axis of symmetry, inducing jetting. Typical examples include bubble bursting at the interface \cite{Duchemin2002,Deike2018,Gordillo2019}, drop impact on a pool~\cite{Oguz1990}, and sequential bubble (droplet) pinch-off by a Rayleigh-Plateau instability. This picture is supported by the absence of bubble entrainment in Fig.~\ref{fig:types}a, where the liquid is more viscous, as viscosity is known to dampen capillary waves \cite{Sanjay2021}. However, simulations with (an unphysical) zero surface tension showed that the process leading to bubble pinch-off is \emph{not} driven by surface tension (see movie SM2 of the supplementary material~\cite{SM}). Similar meniscus oscillations were observed and, although unphysical, pinch-off eventually happens when the protruding meniscus was thinner than the grid size. Thus, it can be concluded that for this system, capillary waves do not govern the meniscus deformations and the resulting pinch-off. Note that that pinch-off can also result from inertial oscillations \cite{bergmann-2009-jfm,gekle-2009-prl,gekle-2010-prl,gekle-2010-jfm,Raja2019,Raja2020}.

\begin{figure}[t]
	\centering
	\includegraphics[width = \textwidth]{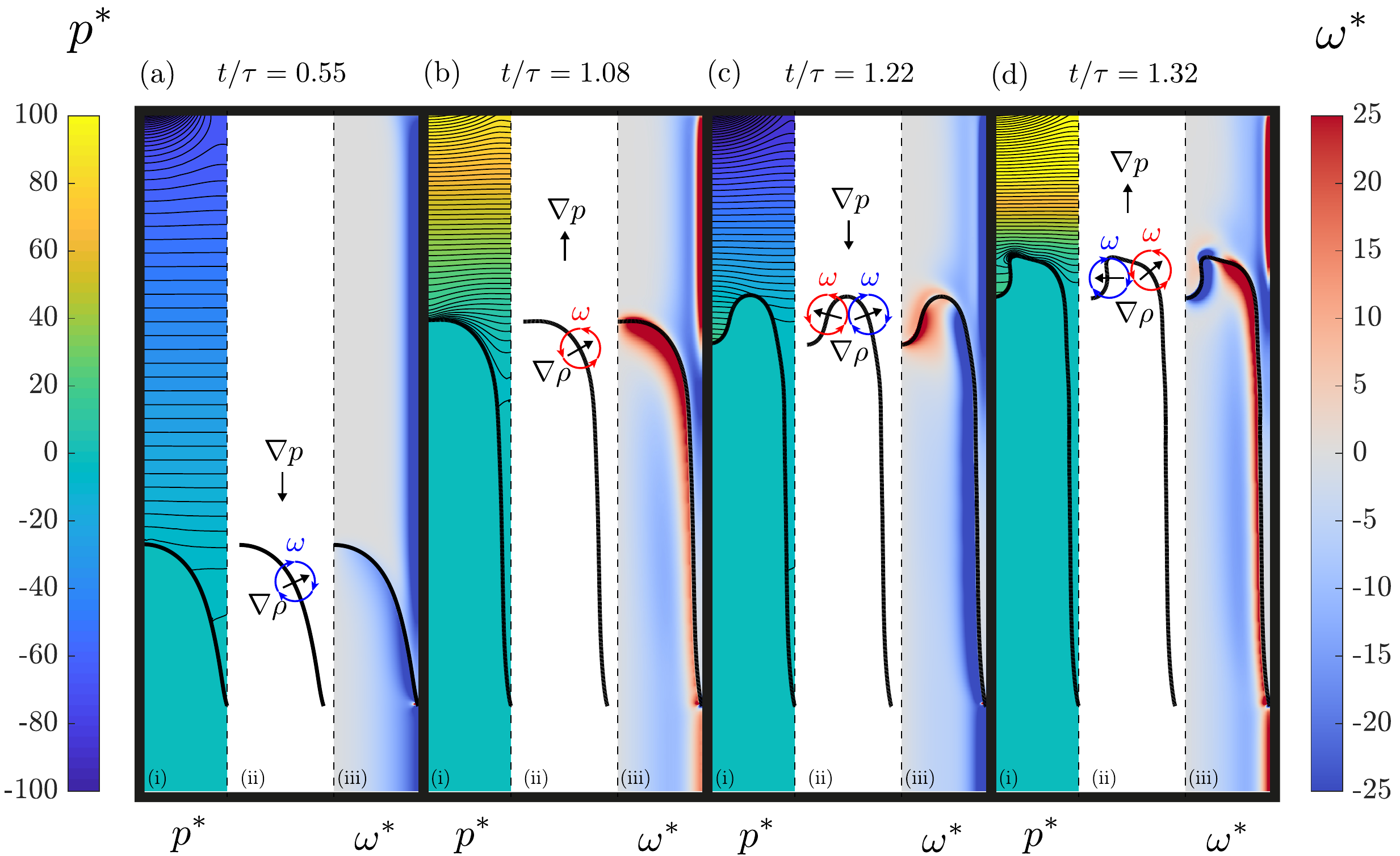}
	\caption{Illustration of the mechanism of meniscus oscillation and bubble entrainment. Four time instants are shown, $t/\tau\in \{0.55,1.08,1.22,1.32\}$. Each time point contains three panels, showing, from left to right, the dimensionless pressure $p^* = p\tau/\left(\rho_lR_n^2\right)$ with isocontours, an explanatory schematic of the baroclinic torque generated at the interface, and the dimensionless vorticity $\omega^* = \omega\tau$, respectively.}
	\label{fig:mechanism}
\end{figure}

To further unravel the mechanism of bubble entrainment, we plot the velocity vector field in Fig.~\ref{fig:vecotrs} (see movie SM3 of the supplementary material~\cite{SM}). Figure~\ref{fig:vecotrs} shows that the flow-focusing mechanism deforms the meniscus, as detailed before in ref.~\cite{Fraters2021Oscillations}. In addition, we now observe well-defined vortical rings that are associated with the flow-focusing at the curved interfaces. These vortical rings alternate in direction. For a deeper understanding of the origin of these vortical rings, we examine the vorticity equation:
\begin{equation}
	\frac{\partial\boldsymbol{\omega}}{\partial t} + \left(\boldsymbol{u}\cdot\nabla\right)\boldsymbol{\omega} = \left(\boldsymbol{\omega}\cdot\nabla\right)\boldsymbol{u} + \frac{1}{\rho^2}\nabla\rho\times\nabla p + \nabla\times\left(\frac{\nabla\cdot\mathcal{D}}{\rho}\right),
	\label{eq::vorticity}
\end{equation}
where $\boldsymbol{\omega} = \nabla\times\boldsymbol{u}$ is the vorticity and $\mathcal{D}$ is the viscous stress tensor. The assumption leading to Eq.~(\ref{eq::vorticity}) is an incompressible flow with conservative body forces. The second term on the right hand side of Eq.~(\ref{eq::vorticity}) is the baroclinic torque, describing the generation of vorticity due to the misalignment of density and pressure gradients. This torque is ubiquitous in geophysical flows \cite{Magnaudet2020}, but is also present in flows with much smaller length scales \cite{Fuster2021,Rossi2021,Wu1995}. This term is particularly relevant in our study since it is responsible for the generation of vorticity at the interface, thus leading to flow-focusing, and eventually to bubble pinch-off. In Fig.~\ref{fig:mechanism}, this mechanism is illustrated by showing four time instants of the numerical simulation. For each time-point, three panels are shown with, from left to right, (i) the dimensionless pressure field with isocontours, (ii) an explanatory schematic, as well as (iii) the dimensionless vorticity. At $t/\tau = 0.55$, the meniscus is retracting inside the nozzle due to a pressure drop near the piezo (top boundary of the domain). The pressure field (i) reflects this behavior and clearly shows the direction of the pressure gradient. In the schematic (ii), the gradient of density is also shown across the interface (arrow inside circle), pointing to the heavier fluid. The baroclinic torque, generated at the interface, is therefore clockwise (blue circular arrow). It must be noted that this vorticity is proportional to $\rho^{-2}$. Therefore, it is much more pronounced in air than in water, as seen in the corresponding vorticity field (iii). At $t/\tau = 1.08$, the meniscus is being pushed outwards. The gradient of pressure thus changes sign, and the generated vorticity at the interface becomes counter-clockwise (red circular arrow). This induces flow-focusing at $r = 0$, the central axis, (Fig.~\ref{fig:vecotrs}a-b), thus contributing to the deformation of the meniscus, into the shape shown at $t/\tau = 1.22$. Another interpretation of this topological change can be the Rayleigh-Taylor instability. And although the latter can be attributed to the baroclinic torque \cite{Roberts2016}, disentangling these two effects is quite difficult if one assumes them to be independent. In appendix \ref{appA}, we discuss this matter further and provide further evidence as to why we think these topological changes are due to the baroclinic torque rather than the exponential growth of a perturbed state. Later on, the meniscus is being retracted. With the current topology, the gradient of density has two different directions, and thus two oppositely rotating vortical rings, focusing the flow as shown in Fig.~\ref{fig:vecotrs}c. At $t/\tau = 1.32$, the meniscus is being pushed again, and so the vortical rings' rotations alternate, focusing the flow at the interface as shown in Fig.~\ref{fig:vecotrs}d. Although the pressure isobars are closely packed at the point of highest curvature in Fig.~\ref{fig:mechanism}d-(i), giving rise to a higher pressure gradient, the flow focusing does not happen there. This observation presents further argument in favor of the vorticity as the flow-focusing mechanism in the present work. A clockwise vortical ring persists at the tip of the protruding meniscus (Fig.~\ref{fig:vecotrs}f), thus contributing to bubble pinch-off, which is determined at the latest stage by capillary forces in a Rayleigh-Plateau like mechanism. If we define a local Weber number based on the magnitude of the vortical ring $We_\omega = \rho\omega^2d^3/\sigma$, where $\omega$ is the magnitude at the centre of the vortical ring and $d$ is the closest distance between this centre and the interface, we get a value of $We_\omega \simeq 50$. This is an indication that the vortical ring is more important than surface tension in leading to the pinch-off of the bubble. Movie SM4 of the supplementary material \cite{SM} bears further evidence to the fact that baroclinic torque is the dominant mechanism, and not surface tension. Movie SM4 depicts a numerical simulation where the densities and the viscosities of the two phases are made equal (artificially). There is still surface tension in the simulation. According to Eq. (\ref{eq::vorticity}), this precludes the generation of any baroclinic torque. Movie SM4 shows that the flow focusing effect and the jet formation are completely absent, despite the presence of surface tension. This indicates that it is indeed the proposed baroclinic torque mechanism that results in the formation of the central air jet.

The last term in Eq.~(\ref{eq::vorticity}), which can be written as $\nu\nabla^2\omega$, $\nu$ being the kinematic viscosity, is a diffusion term. Therefore, as the viscosity of the liquid increases,  the vorticity generated at the interface gets diffuses into the bulk of the fluid. Hence, its flow focusing effect decreases, which can support the observation of no bubble pinch-off in Fig.~\ref{fig:types}d.

Since boundary integral simulations assume an irrotational flow, vortices are a missing ingredient in the bulk of the liquid; however, vorticity still exists at the interface \cite{baker-1984-physicad}. Fraters \textit{et al.} (2021)~\cite{Fraters2021Oscillations} still obtained a proof-of-concept of the flow focusing mechanism leading to pinch-off by imposing an \textit{ad-hoc} boundary condition on the velocity. Also, in their work, the contact line was not pinned and the interface was initialized with a certain curvature that was optimized to capture the experimental observations. We attribute the flow-focusing in their BI simulations to pressure impulses due to the abrupt changes in the imposed velocity field, similar to the interpretation by Peters \textit{et al.} (2013)~\cite{Peters2013}. That process is therefore similar to the flow-focusing induced on the curved meniscus in the capillary tube impact experiments performed by Antkowiak \textit{et al.} (2007)~\cite{Antkowiak2007}. However, the current DNS clearly show that the flow-focusing, leading to bubble pinch-off from an inkjet meniscus, is vorticity-induced rather than being the result of pressure impulses that induce flow-focusing on the curved interfaces. Therefore, the current work sheds more light on the flow-focusing mechanism, and considers an experimental configuration with realistic boundary conditions, unlike the previous work which merely presented a proof-of-concept. And thus, the present work does not nullify its predecessor but rather complements it.

\section{Conclusion}\label{sec:Conclusion}

The oscillating flows in an inkjet printhead can lead to geometric flow-focusing through which an inward gas jet can form on the retracted concave meniscus. The gas jet can pinch-off to entrain an air bubble. The study of the bubble pinch-off phenomena using ultrafast X-ray phase-contrast imaging provided a complete visualization of the oscillating meniscus, which was input to direct numerical simulations. The numerical simulations demonstrated that the presence of vortical rings leads to geometric flow-focusing, which in turn can result in bubble pinch-off. The vortical rings in this system are a manifestation of the baroclinic torque at the interface. The viscosity of the liquid influences the diffusion of vorticity from the interface into the bulk. Therefore, increasing the viscosity will also reduce the probability of bubble entrainment. These results provide a fundamental understanding of the oscillating meniscus and the resulting flows inside an inkjet printhead that allow for a more complete understanding of the bubble entrainment process in an inkjet nozzle.

\section*{Acknowledgments}
The authors would like to thank Kirsten Harth her assistance in the application for the synchrotron and for experimental assistance during the beam time, and Javier Rodr\'{\i}guez-Rodr\'{\i}guez for insightful discussion.  This work was supported by an Industrial Partnership Programme of the Netherlands Organisation for Scientific Research (NWO), co-financed by Canon Production Printing Netherlands B.V., University of Twente, and Eindhoven University of Technology. This research used resources of the Advanced Photon Source, a U.S. Department of Energy (DOE) Office of Science user facility at Argonne National Laboratory and is based on research supported by the U.S. DOE Office of Science-Basic Energy Sciences, under Contract No. DE-AC02-06CH11357.

\appendix
\section{Rayleigh-Taylor instability}
\label{appA}
\begin{figure}[t]
	\centering
	\includegraphics[width = 0.5\textwidth]{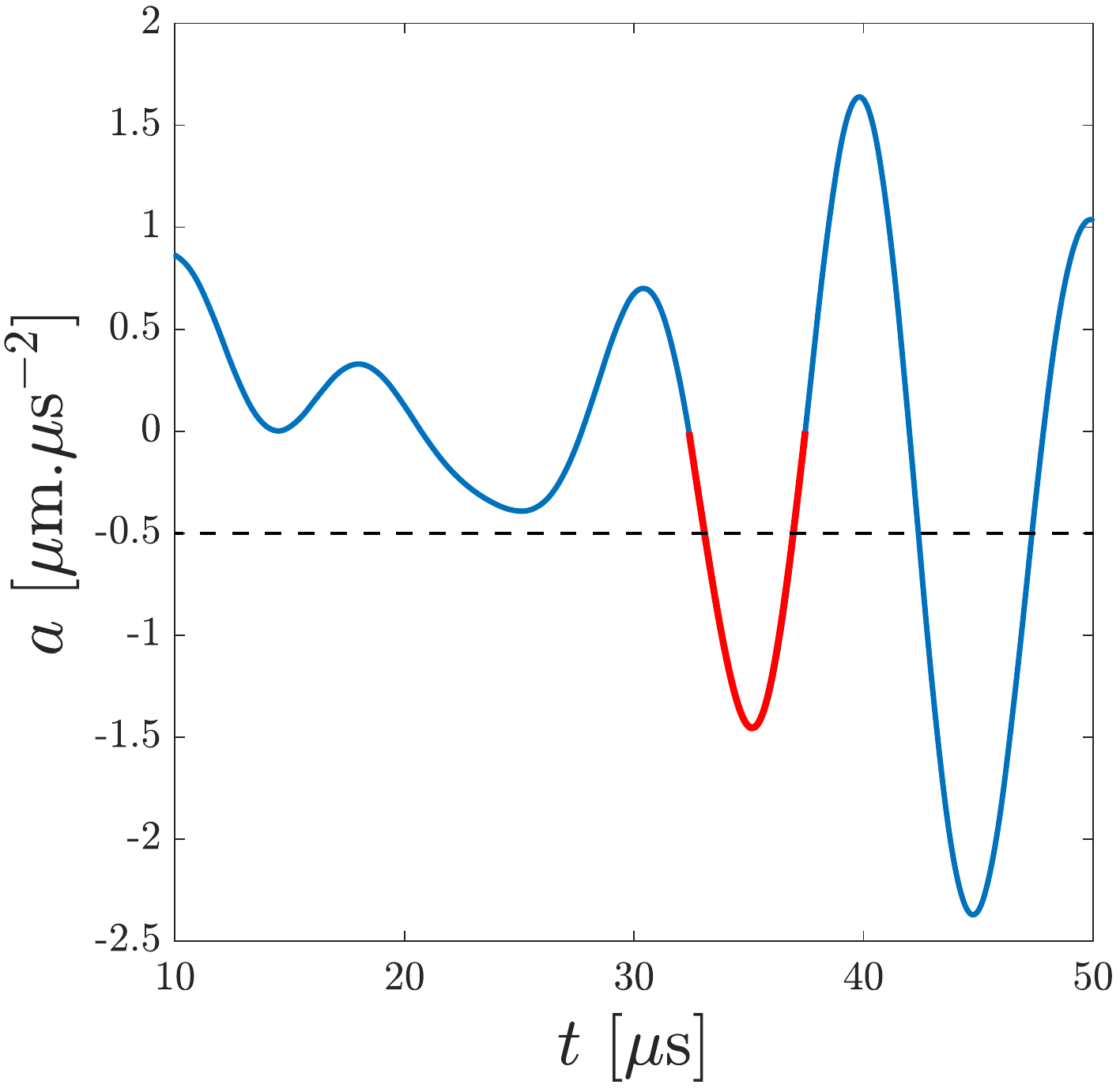}
	\caption{Average acceleration of the liquid inside the nozzle with respect to time. The dashed line indicates the approximate average of the deceleration phase responsible of the topological change in figure \ref{fig:mechanism}c.}
	\label{fig:appendix1}
\end{figure}
An accelerated interface between two fluids of different densities is prone to an instability if the gradient of density is opposite to the acceleration \cite{Rayleigh1882,Taylor1950}. A perturbation with a given wavelength will then grow exponentially. Such a configuration is unstable for all wavelengths provided the acceleration is constant, or vice-versa. However, if surface tension is present, this ceases to be true since capillary forces will work as a stabilizing agent, and would flatten the interfacial corrugations, mainly those with a small wavelength. In other words, if surface tension is taken into account, there exits a critical wavelength/acceleration below which the perturbation is stable for a given acceleration/wavelength. By performing a linear stability analysis \cite{Chandrasekhar2013}, it is found that the critical acceleration $a_c$ needed to destabilize a flat interface with an infinitesimal perturbation of wavelength $\lambda$ is
\begin{equation}
	a_c = -\frac{4\pi^2\sigma}{\lambda^2|\Delta\rho|}.
	\label{eq::RTc}
\end{equation}
In our case, Fig. \ref{fig:mechanism}b shows the interface before the topology change is observed in Fig. \ref{fig:mechanism}c, and that might be attributed to a Rayleigh-Taylor instability. Although the interface is not flat, we shall apply Eq. (\ref{eq::RTc}) to get the critical acceleration above which our interface becomes unstable. From figure \ref{fig:mechanism}c, we estimate a wavelength of the order of the nozzle radius $\lambda\simeq R_n$. Equation (\ref{eq::RTc}) then yields a critical acceleration $a_c = -2.5~\mu$m.$\mu$s$^{-2}$.

Figure \ref{fig:appendix1} shows the average acceleration of liquid water inside the nozzle with respect to time. It is derived from the boundary condition on the flow rate shown in Fig. \ref{fig:extracting_flowrate}d. The deceleration phase responsible of the topological change seen in Fig. \ref{fig:mechanism}c is coloured in red. Its average is $\bar{a}\simeq -0.5~\mu$m.$\mu$s$^{-2}$, which is an order of magnitude smaller than the critical acceleration $a_c$. However, this is an inconclusive evidence since the analysis might not be readily applied to the configuration at hand.

\begin{figure}[t]
	\centering
	\includegraphics[width = \textwidth]{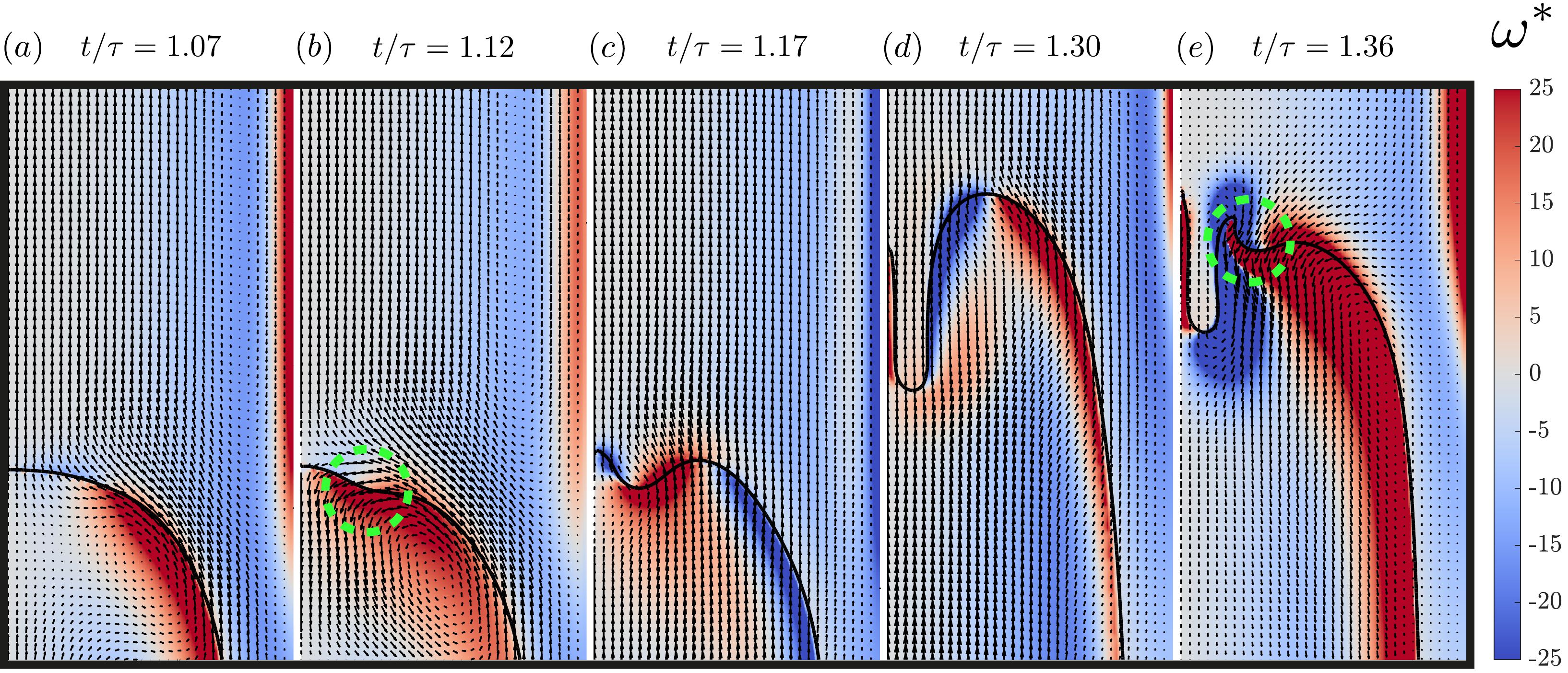}
	\caption{Snapshots from numerical simulations of the meniscus oscillations leading to bubble entrainment. The velocity is depicted as a vector field, and the dimensionless vorticity $\omega^* = \omega\tau$ is illustrated in the background as a color code. In each snapshot, the interface is drawn in black. In this simulation, surface tension is artificially set to zero. The green dashed circles indicate the location of flow focusing at the interface due to the baroclinic torque.}
	\label{fig:appendix2}
\end{figure}

We present further evidence as to why we think these topological changes are due to the baroclinic torque rather than to a Rayleigh-Taylor instability. Figure \ref{fig:appendix2} depicts a sequence of snapshots from the numerical simulation where surface tension is set to zero (see movie SM2 of the supplementary material~\cite{SM}). Since capillary forces are not taken into account, any deceleration will make interfacial perturbations RT unstable. It must be stated that the deceleration is maximum at the axis of symmetry, therefore any Rayleigh-Taylor instability should appear at $r = 0$. However, this is not the case as can clearly be seen from Fig. \ref{fig:appendix2}b. The flow is focused at the location indicated by the dashed green circle, due the to the vorticity in the liquid, especially since the interface provides no resistance due to the lack of surface tension. Subsequently, this topological change develops to the shape in Fig. \ref{fig:appendix2}d. Then again, one clearly sees the flow focusing at the location indicated by the dashed green circle in Fig. \ref{fig:appendix2}e. It is clear from the oblique deformation of the interface that it is due to the baroclinic torque generated at the interface. As previously mentioned in the manuscript, this also rules out capillary waves as the main driver of this mechanism.

\bibliography{bibliography}

\begin{thebibliography}{77}
\bibitem [{SM()}]{SM}%
  \BibitemOpen
  \href@noop {} {\enquote {\bibinfo {title} {See {S}upplemental {M}aterial at
  ({URL} will be inserted by publisher) for the movie corresponding to the
  snapshots in fig.~\ref{fig:setup}d.}}\ }\BibitemShut {NoStop}%
\end{thebibliography}

\end{document}